\documentclass{article}

\usepackage{arxiv}

\usepackage[utf8]{inputenc} 
\usepackage[T1]{fontenc}    
\usepackage{hyperref}       
\usepackage{url}            
\usepackage{booktabs}       
\usepackage{amsfonts}       
\usepackage{nicefrac}       
\usepackage{microtype}      
\usepackage{lipsum}
\usepackage{graphicx}
\usepackage[ruled,linesnumbered]{algorithm2e}
\usepackage{amsmath}
\usepackage{url}
\usepackage{setspace}
\graphicspath{ {./images/} }

\SetKwInput{KwInput}{Input}               
\SetKwInput{KwOutput}{Output}

\title{Utilizing Deep Learning to Identify Drug Use on Twitter Data}

\author{
 Joseph Tassone \\
  Department of Computer Science \\
  Lakehead University \\
  Thunder Bay, ON, P7B 5E1 \\
  \texttt{jtasson2@lakeheadu.com} \\
  \And
 Peizhi Yan \\
  Department of Computer Science \\
  Lakehead University \\
  Thunder Bay, ON, P7B 5E1 \\
  \And
 Mackenzie Simpson \\
  Department of Computer Science \\
  Lakehead University \\
  Thunder Bay, ON, P7B 5E1 \\
  \And
 Chetan Mendhe \\
  Department of Computer Science \\
  Lakehead University \\
  Thunder Bay, ON, P7B 5E1 \\
  \And
 Vijay Mago \\
  Department of Computer Science \\
  Lakehead University \\
  Thunder Bay, ON, P7B 5E1 \\
  \And
 Salimur Choudhury \\
  Department of Computer Science \\
  Lakehead University \\
  Thunder Bay, ON, P7B 5E1 \\
}

\begin{document}
\maketitle
\begin{abstract}
The collection and examination of social media has become a useful mechanism for studying the mental activity and behavior tendencies of users. Through the analysis of collected Twitter data, models were developed for classifying drug-related tweets. Using topic pertaining keywords, such as slang and methods of drug consumption, a set of tweets was generated. Potential candidates were then preprocessed resulting in a dataset of 3,696,150 rows. The classification power of multiple methods was compared including support vector machines (SVM), XGBoost, and convolutional neural network (CNN) based classifiers. Rather than simple feature or attribute analysis, a deep learning approach was implemented to screen and analyze the tweets' semantic meaning. The two CNN-based classifiers presented the best result when compared against other methodologies. The first was trained with 2,661 manually labeled samples, while the other included synthetically generated tweets culminating in 12,142 samples. The accuracy scores were 76.35\% and 82.31\%, with an AUC of 0.90 and 0.91. Additionally, association rule mining showed that commonly mentioned drugs had a level of correspondence with frequently used illicit substances, proving the practical usefulness of the system. Lastly, the synthetically generated set provided increased scores, improving the classification capability and proving the worth of this methodology.
\end{abstract}

\keywords{NLP \and CNN \and Twitter Analysis \and Big Data}

\section{Introduction}
\label{sec:introduction}
Collecting accurate and up-to-date trend information regarding drug-use is an arduous task \cite{sources_of_error}. The illicit nature of the topic makes surveying a population difficult, as the potentially illegal nature tends to generate a less honest or unwilling response. This limits the usefulness of the data collected and provides a demand for an accurate system. A prospective solution is in social media, which has been used as a source for studying the mental activity and behavior tendencies of users \cite{sarker2016social}. Current research has gone so far as to suggest the possible validity in utilizing the information posted online as a substitution for actual surveyed data \cite{gittelman2015new,kim2017classification}. This fact is not necessarily surprising, as there is widespread utilization and sites such as Twitter are consistently accessed by a significant population of people. As social media is prevalent in today's society, it provides an excellent opportunity for developing a generalized drug detection system, as well as a manner for extracting relevant trends.

Twitter data is not the most consistent or stable information to work with\cite{du2018extracting}. Inconsistencies within the wording and the lack of discrete variables made analysis and classification a difficult task. Traditional machine learning methods have proved ineffective for our purposes (see Section \ref{results}). As a result, a deep learning approach was used in this research to screen and analyze positively referenced, drug-related tweets. Topic pertaining keywords, such as slang and use-conditions (methods of drug consumption) were used to collect Twitter data. A subset of dataset was then manually labeled with two categories: positive or negative. For clarity, a text such as ``smoke weed every day" would register as drug-positive, while a tweet like ``all drugs should be illegal" would register as drug-negative. A normal text without reference to any particular drug would also register with a negative result. Following this, a deep learning model using a convolutional neural network (CNN) \cite{kim2014convolutional} was trained on the labeled data to classify between positive and negative. A {\it word2vec} algorithm was used which allowed the embedding of alike words (words having a comparable meaning) to be seen as similar \cite{mikolov2013efficient}. This helped the CNN care less about the variety of words and instead focus on the semantic meaning of words and their corresponding relationships. Further details are provided in Section \ref{methodology}.

Training the CNN showed a strong capability for accurately classifying, the details of which can be found in Section \ref{cnn_results}). Additional classification methods were used to compute the quality of classification; however, the deep learning algorithm with synthetic data was found to outperform them (see Section \ref{results}). Deep learning is not a unique method for performing this task; however, there is little research in utilizing it as a general drug detection system. At the same time, our research found promising results in combining this methodology with synthetically generated data (see Section \ref{data_preprocess}). The following points have been concluded from this research:

\begin{itemize}
  \item[$\bullet$] This work verifies that the ``possibly\_sensitive" tag generated by the Twitter API cannot be used for the classification of drug-positive tweets.
  \item[$\bullet$] A CNN model with synthetic data was developed and outperformed other methods in classifying drug-related tweets.
  \item[$\bullet$] A novel approach of generating labeled synthetic data improved the accuracy and classifying capability of the model.
  \item[$\bullet$] Commonly mentioned drugs had a level of correspondence with frequently used illicit substances.
\end{itemize}

The remainder of this paper is arranged as follows. Section \ref{related_work} describes the related work, emphasizing previous or similarly documented techniques. Section \ref{methodology} presents the methodology utilized, with details on the data and pre-processing performed and the quality of the data. Section \ref{results} shows the experimental results of the SVM, XGBoost, and CNN-based classifiers. Additionally, respective keyword strength and patterns in the data were determined. Section \ref{conclusion} concludes the paper and discusses possible future works.

\section{Related Work}
\label{related_work}
\subsection{Health Analysis using Social Media Data}
Social media data reflects a population's characteristics, including public health information. Many social media platforms such as Twitter and Facebook have a massive user base, constantly generating an enormous amount of messages. For instance, based on a 2019 statistic, 500 million tweets were sent out on a daily basis. Therefore, monitoring and analyzing social media data should be prominent in population-based research, including public health.

In \cite{lampos2010flu}, Lampos et al. proposed a way to detect and track influenza in the United Kingdom. Their method was utilizing regression in learning a set of weighted keywords to compute a score, which reflected the influenza rate. Paul et al. \cite{paul2011you} assumed that each health-related tweet reflected an underlying ailment and proposed an ailment topic aspect model (ATAM) for syndromic surveillance. The results showed the broad applicability of analyzing Twitter data for public health research. The authors also pointed out the limitation of using Twitter data due to the age of users (Twitter users tend to be teenagers or young adults). Chew et al. leveraged Twitter data analysis to track the trend of sentiment and public attention during the 2009 H1N1 pandemic \cite{chew2010pandemics}. Besides the text content and the meta-information, such as keywords count, \cite{gittelman2015new} introduced a predictive model for the classification of healthy and unhealthy populations based on Facebook ``likes". This work also showed that the significant value of Facebook ``likes" in public health prediction and population health-related behaviour analysis. As mentioned above, these researches were focused on population-level health status rather than the study of individual users.

Analyzing individual-level health status of social media users helps doctors or healthcare professionals detect potential patients and provide help. In \cite{heaivilin2011public}, Twitter data was used for dental pain surveillance. Since dental pain is non-infectious, the purpose of their research was to detect Twitter users with a toothache via data mining. Researchers developed a coding system to analyze the content of the collected tweets. Other similar works include, Coppersmith et al. who built a binary classifier to detect the post traumatic stress disorder of individual Twitter users \cite{coppersmith2014measuring}.

\subsection{Drug abuse detection in social media}
Due to the prevalence of social media, research on detecting and monitoring drug abuse-related behaviors have been carried out in recent years. The methods used in these researches can be categorized into traditional statistic methods and machine learning approaches. A semantic prescription drug abuse surveillance platform (PREDOSE) was introduced in \cite{cameron2013predose}. The study scope of the semantic data of PREDOSE is web forum posts. PREDOSE only dealt with three types of data: entities, relationships, and semantic triples. The first stage of PREDOSE was to collect and clean the posts. In the second stage, domain knowledge in drug abuse studies was leveraged to extract and process the information of interest. In the third stage, statistic-based qualitative and quantitative analysis was used to detect the drug user attitudes and behaviors; while temporal analysis was applied to detect the trend of drug abuse.

Sarker et al. proposed a hybrid classification model for automatically monitoring prescription medication abuse from Twitter data \cite{sarker2016social}. The hybrid classification model was a combination of four traditional supervised learning algorithms, namely: Na\"{i}ve Bayes, support vector machine, maximum entropy, and a decision tree-based classifier. Since the distribution of abuse and non-abuse tweets was highly unbalanced, the resulting model had a high accuracy yet a poor F1 score. 

In \cite{kim2017classification}, the researchers collected tweets with E-cigarette related keywords and manually annotated a small set of data for analysis. The annotated data had five categories, representing the type of corresponding user: individuals, vaper enthusiasts, informed agencies, marketers, and spammers. The classifier used in this work was gradient boosting regression trees. They further studied the importance of each feature regarding the user types. There are some limitations within this work, such as manual feature engineering and relatively small training dataset.

Social media data contains a great deal of metadata, such as a user's basic information and their interpersonal relationship network. As such, analyzing the high dimensional patterns within this could help enhance the user classification accuracy. In \cite{kursuncu2018s}, Kursuncu et al. leveraged three levels of features (person-level, content-level, and network-level) in Twitter data for representing a user, where each level of features was called a view. Compositional multi-view embedding (CME) was used for embedding the three levels of features. Experimental results showed that the classification accuracy was improved by using CME.

Hu et al. proposed a deep learning-based Twitter posts drug abuse risk behavior detection system \cite{huang2018deep}. In their approach, a small number of labeled tweets was used for training the CNN classifier. They further used the CNN to label some of the unlabeled tweets to augment the training dataset. By repeating the above-mentioned steps, the classification accuracy was improved. The problem was that the approach might reinforce the ability of the CNN classifier to detect the patterns of the original manually labeled data, yet miss other patterns which are not in the original labeled data.

\subsection{Social Media Text Analysis with Deep Learning}
Social media data is worthwhile to mine, as people nowadays tend to express their thoughts through social network platforms \cite{serrat2017social}. Du et al.\cite{du2018extracting} proposed a deep learning approach to extract psychiatric stressors for suicide from twitter data. Keyword-based querying and filtering was used to screen the possible suicide-related tweets from the collected Twitter stream. Following this, a small subset of the candidate tweets was manually labeled (positive/negative) and trained on a CNN. The model was utilized to further select some suicide-related tweets from the candidates. Finally, training with a RNN was completed to perform the stressor recognition task.

Sawhney et al.\cite{sawhney2018exploring} proposed a long short-term memory recurrent neural network (RNN) to classify suicidal ideation-related social media sentences. They used a two-level embedding approach to prepare the input data for the RNN. The first level of embedding was a 300-dimensional word2vec embedding, while the second level was a sentence level embedding; where a single-layer CNN was used to generate feature maps. They concatenated the pooled feature maps relative to the order of words in the input sentence and used an RNN to do the final classification. The sentence level CNN embedder and the RNN were optimized during training. Severyn and Moschitti used a CNN in Twitter sentiment analysis \cite{severyn2015twitter}. As the parameters in a CNN are randomly initialized before training and a proper initialization of parameter values is crucial to train a good model; they used a pre-training method as an initialization approach. This was done prior to training the model on their target training dataset. Their pre-training dataset was a set of ten million tweets containing positive words. The main drawback of their approach was that the pre-training process took a significant amount of time (a few days).


\section{Methods}
\label{methodology}
\subsection{Data Source}
We created a Twitter developer account which allowed us to employ Twitter's data infrastructure tools and utilize the collected information for research. All the techniques and data mentioned conforms to the Developer's Agreement and Policy enforced by the organization \cite{twitterdevagree}. The data consists of extracted social media information, obtained through Twitter's official public API. Specifically, it is a collection of tweets that were pulled based on 157 keywords; all of which were related to either specific drugs or drug-uses. Drug-use keywords included: ``snorted", ``snorting", ``snort", ``pills", ``blotter paper", ``blotting paper", ``tabs", ``patches", ``injecting", ``injected", ``inject", ``ingesting", ``ingested", ``ingest", ``smoked", ``smoking", ``smoke", ``chewed", ``chewing", ``chew", ``vaporized", ``vaporizing", ``vaporize", ``vaped", ``vaping", ``vape", ``bong", ``pipe", ``joint", ``needle", ``shoot up", ``hookah", ``grinder", ``one hitter", ``sinker", ``popper", ``inhaling", ``inhaled", and ``inhale" \cite{deauseterms}. The remaining drug-only keywords can be viewed in Table~\ref{drug_words}. These keywords were chosen based on an intelligence report published by the Drug Enforcement Agency (DEA), categorizing drugs by slang and street terms \cite{deaslangterms}.

\begin{table}[h!]
\centering
\caption{Drug Related Keywords.}
\label{drug_words}
\setlength{\tabcolsep}{3pt}
\resizebox{0.5\textwidth}{!}{%
\begin{tabular}{|l|l|}\hline
\textbf{Drug Category} & \textbf{Keyword}  \\ \hline
Amphetamine            & amy, bennies, benz, dexies, diet pills, \\&get ups, pep pills, wake-ups, amphetamine                       \\ \hline
Cocaine                & blow, coke, crack, nose candy, cocaine                         \\ \hline
DMT                    & dimitri, dmt       \\ \hline
General                & drugs, drug        \\ \hline
GHB                    & georgia home boy, grievous bodily harm, \\&ghb, liquid ecstasy, liquid e, liquid x
\\ \hline
Heroin                 & black tar, brown sugar, china white, heroin,\\& mexican brown, skag, white horse
\\ \hline
Hydrocodone            & 357s, dro, fluff, norco, vics, vikes, watsons, \\&hydrocodone    \\ \hline
Ketamine               & cat valium, special k, vitamin k, ketamine                      \\ \hline
Klonopin               & k-pin, super valium, klonopin                            \\ \hline
LSD                    & acid, blotter acid, blotter, electric kool aid,\\& lucy in the sky with diamonds, microdot, \\&tabs, lsd          
\\ \hline
Marijuana              & 420, blunt, bud, dagga, dope, ganja, grass, \\&green, hashish, hash, hemp, herb, mary jane, \\&pot, weed, marijuana                                   \\ \hline
MDMA                   & e, ecstasy, happy pill, love drug, molly,\\& vitamin e, xtc, mdma \\ \hline
Mescaline              & blue caps, media luna, mescal, mezcakuba, \\&topi, mescaline  
\\ \hline
Methamphetamine        & crank, crystal, meth, shards, speed, tweak,\\& uppers, methamphetamine                            \\ \hline
Mushrooms              & boomers, baps, mushies, shrooms, tweezes, \\&mushrooms        \\ \hline
Nitrous\_Oxide         & buzz bomb, laughing gas, nitrous, nox, \\&whippets, nitrous oxide  \\ \hline
Opioid                 & abstral, acetaminophen, actiq, china girl, \\&codeine, dance fever, dilaudid, duragesic, \\&exalgo, fentanyl, hydromorphone, lorcet, \\&lortab, methadone, morphine, murder 8, \\&onsolis, oxy, oxycodone, oxyContin, \\&oxymorphone, percocet, vicodin, opioid 
\\ \hline
PCP                    & angel dust, love boat, peace pill, superweed, \\&pcp            \\ \hline
Peyote                 & black button, green button, hikuli, hyatari, \\&peyote             \\ \hline
Ritalin                & ritalin            \\ \hline
Steroids               & gym candy, pumpers, roids, steroids                             \\ \hline
Synthetic\_Cathinones  & bath salts, bloom, cloud 9, cloud nine, \\&cosmic blast, flakka, ivory wave, lunar wave, \\&vanilla sky, white lightning, \\&synthetic cathinones             \\ \hline
Xanax                  & benzos, xanies, z bars, zanbars, xanax      \\ \hline      
\end{tabular}
}
\end{table}

Twitter data collection ran from October 22 to November 30, 2018. Only tweets containing the valid keywords were selected, and mispellings were handled on a case-by-case basis by the Twitter API. The initial set was cleaned with the following filters: removed newlines, contracted extra spaces, removed hashtags, removed emojis, removed reserved words, removed smiley, removed URLs, removed mentions, removed all punctuation, remove all numbers, converted the text to lowercase, removed stop words, fixed known misspellings, and contracted words. This cleaned dataset consisted of 51 attributes, with 26,184,358 tuples of data (see Additional Files) and is available for research purposes upon request.

\subsection{Data Preprocessing}
\label{data_preprocess}
\begin{figure}[h!]
  \centering
  \includegraphics[width=0.95\textwidth]{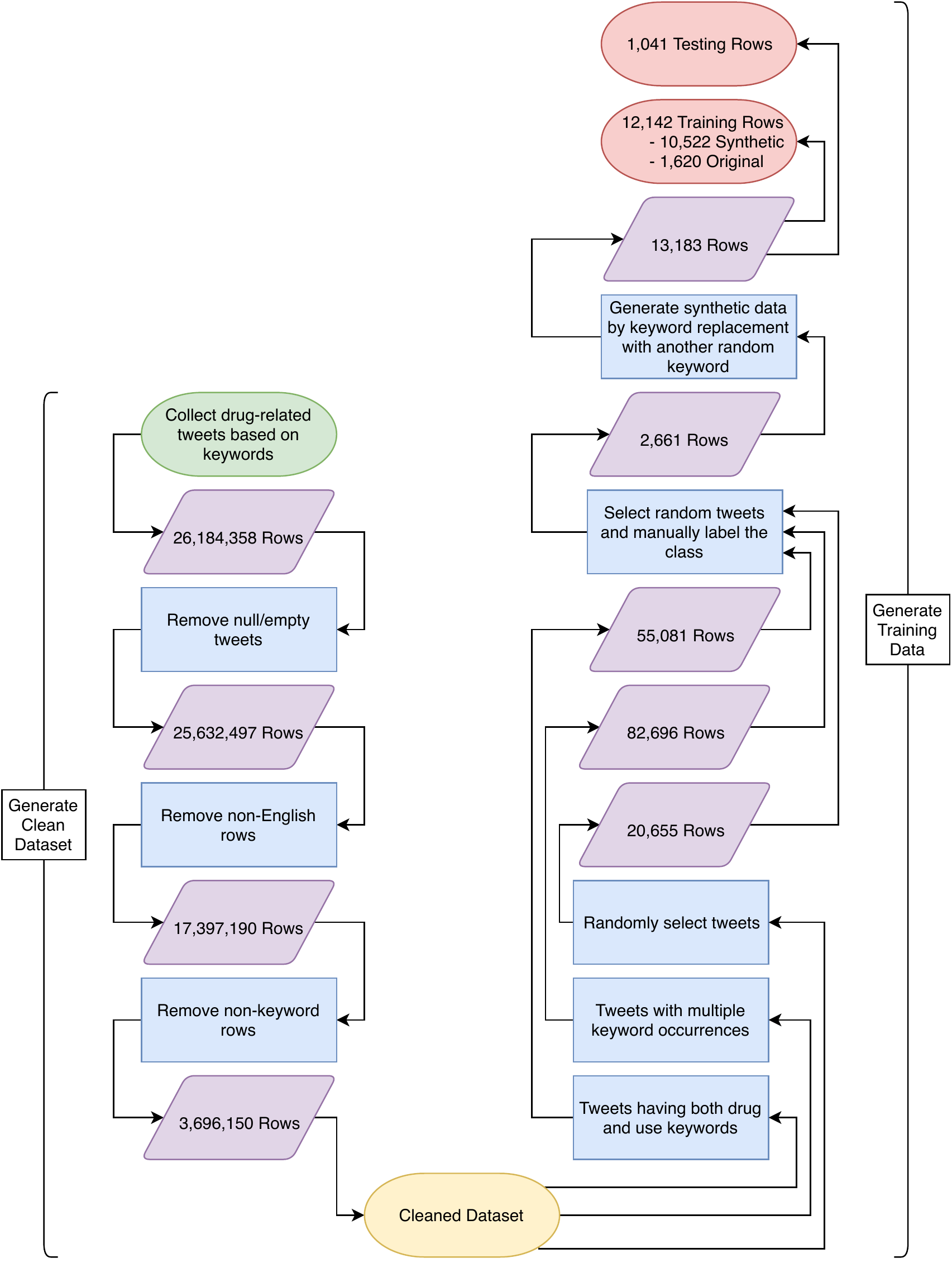}
  \caption{Preprocessing steps for each stage of the collected data. Circles represent filters placed on the data, while rectangles represent the updated set size.}
  \label{preprocess}
\end{figure}

The organization of this data demanded a large amount of pre-processing. The tweets that were received were not tagged according to drug-use; therefore, this needed to be determined prior to analysis. In addition, by the sheer volume of the tweets alone, tagging by hand was not a realistic option. As a result, a CNN was trained with a subset of the data to perform the remainder of the task. Many of the tweets also proved to be irrelevant for the purposes of training the neural network; therefore, were removed. The following filters were utilized in generating the dataset following initial collection (see Figure~\ref{preprocess}):

\begin{itemize}
\item[$\bullet$] Removed rows with null or empty tweets.
\item[$\bullet$] Removed non-English rows.
\item[$\bullet$] Removed rows with tweets having no keywords.
\end{itemize}

For the purposes of data analysis, only English tweets were considered, as other languages would have added a level of complexity and required translation. There was the possibility after the initial cleaning that there would either blank tweets or those without keywords. An example of this would be if a keyword was held in a hashtag or a URL. After these filters were considered, the resulting set consisted of 3,696,150 rows. This may seem like a heavy reduction; however, the filtering procedure is strict in order to ensure data quality. At this stage, a row in the dataset consisted of each collected tweet and the associated metadata attributes attached to it.

The previously mentioned pre-processing tasks standardized the dataset, and initially an attempt at simplifying the information by reducing the number of keywords. The {\it method of delivery} for a drug was replaced by the literal word ``BETA", while the specific drug was replaced by ``ALPHA". As an example, a phrase like ``Smoke weed everyday" would become ``BETA ALPHA everyday". It was theorized that the neural network did not need to interpret the drug that the user was referring to, and the name of the substance could be replaced with a pseudonym. While it was thought this would reduce the burden of training the model, it resulted in a loss of descriptive intent and decreased the model's accuracy. Furthermore, the data was initially extracted using 290 keywords and was later adjusted to 157. The reasoning behind this reduction was that uncommon slang words like ``friend" (for fentanyl) significantly impacted the training performed on the network. Essentially these words were too common in normal speech or in the case of a word like ``amp" (amphetamine), was detected as the ASCII characterization of ``\&". This resulted in a massive number of drug-negative tweets, risking a skewing in the final metrics. Although it would appear a significant amount of data was removed, these words were uncommon slang terms and the most prevalent keywords still remained. 

During the pre-processing task, additional attributes were generated (example: ``number of keywords used") from the text string for possible variable selection and final analysis. Some of these were in addition to the metadata attributes that were collected along with the tweet by the Twitter API. The attributes following pre-processing (either sums or identifiers) included: ``id\_str", ``text", ``user\_followers\_count", ``possibly\_sensitive", ``timestamp\_ms", ``lang", ``original\_text", ``user\_friends\_count", ``alpha", ``beta", ``snort", ``blotter", ``inject", ``ingest", ``smoke", ``chew", ``vaporize", ``vape", ``inhale", ``hitter", ``shoot", ``tabs", ``patches", ``pills", ``bong", ``pipe", ``joint", ``needle", ``hookah", ``grinder", ``sinker", ``popper", ``Amphetamine", ``Cocaine", ``DMT", ``General", ``GHB", ``Heroin", ``Hydrocodone", ``Ketamine", ``Klonopin", ``LSD", ``Marijuana", ``MDMA", ``Mescaline", ``Methamphetamine", ``Mushrooms", ``Nitrous\_Oxide", ``Opioid", ``PCP", ``Peyote", ``Ritalin", ``Steroids", ``Synthetic\_Cathinones", ``Xanax", ``both", and ``classification". These attributes together made up each row of the dataset.

 The ``classification" attribute required the CNN to be trained, meaning testing and training sets needed to be extracted from the pre-processed dataset. Three temporary sets were generated, based either on patterns or a random selection of data:

\begin{itemize}
\item[$\bullet$]\textbf{Set 1:} Tweets containing both drug and use-keywords.
\item[$\bullet$]\textbf{Set 2:} Tweets containing multiple occurrences of keywords (example: ``weed" is mentioned twice).
\item[$\bullet$]\textbf{Set 3:} Tweets randomly selected from the cleaned 3,696,150 rows.
\end{itemize}

These sets were completely unique, with no overlapping data between them. Following this, 2,661 rows were randomly selected from the set for manual labeling by the Lakehead University DaTaLab students (Mannila Sandhu and Tanvi Barot). Each tweet was assigned either a 1 (drug-positive) or 0 (drug-negative), depending on the semantic meaning of the text. A text such as ``smoke weed every day" would register as drug-positive, while a tweet like ``all drugs should be illegal" would register as drug-negative. Passive references such as those related to news or simple drug discussion would also register as drug-negative. Likewise, news related tweets or those referring to the observation of drug use would be considered drug-negative. Essentially, the tweet had to be referring to the active usage of drugs with an implied or directly supportive connotation in order to garner a positive label. The labelled tweets were then verified by social work student Caleb Pears (specializing in addictions research) to ensure all the classifications were consistent. 

\begin{figure}[h!]
  \centering
  \includegraphics[width=0.8\linewidth]{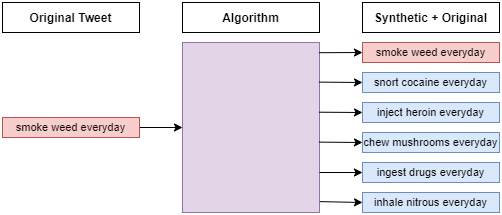}
  \caption{Synthetic data generation process.}
  \label{fig:synthetic_generation}
\end{figure}

This dataset was limited, as the small size meant there was less coverage among the keywords. Training the CNN with this type of data could have allowed a higher probability of misclassification. To ensure proper inclusion and maintain the semantic meaning of the text, synthetic data was generated to compensate. The full process is described through Algorithm \ref{synthetic-algorithm} and Figure \ref{fig:synthetic_generation}, with a specific example in Table \ref{tab:synthetic_generation}. This was done to minimize the sampling bias that could have been present in the neural network. The algorithm simply functioned by looping through the set of tweets, and another loop went through each respective tweet, searching for keywords and replacing them with a random of the corresponding type. Keywords within the ``text" attribute of the 2,661 rows were replaced respectively by either a random drug (from the 157 mentioned in Table~\ref{drug_words}) or drug-use keyword. 13,183 rows were contained in this set, among which 12,142 were allotted for training and 1,041 for testing. As per the Additional Files section, these datasets are available for research purposes upon request. A summary of this methodology, as well as the initial steps, can be seen in  Figure~\ref{preprocess}.

\begin{algorithm}[h!]
 \KwInput{The set of tweets: $T$; the set of use-keywords: $U$; the set of drug-keywords: $V$}
 \KwOutput{A synthetic set of tweets: $W$}
 $n \leftarrow length(T)$\;
 \For{$i \gets 1$ to $n$}   {
    $t\leftarrow$ $i$ tweet of $T$\;
    $m\leftarrow$ number of synthetic tweets to generate\; 
    \For{$j \gets 1$ to $m$}   {
        $w\leftarrow$ $t$\;
        $w'\leftarrow$ Replace use-keyword from $w$ with another random use-keyword\; 
        $w''\leftarrow$ Replace drug-keyword from $w'$, with another random drug-keyword\;
        Add $w''$ to $W$\;
    }
 }
\caption{Synthetic training data generation.}
\label{synthetic-algorithm}
\end{algorithm}

\begin{table}[h!]
\centering
\setlength{\tabcolsep}{3pt}
\caption{Example of synthetic data generated from an original tweet.}
\label{tab:synthetic_generation}
\resizebox{0.6\textwidth}{!}{%
\begin{tabular}{|l|l|l|l|}
\hline
\textbf{Tweet}          & \textbf{Use-Keyword} & \textbf{Drug-Keyword} & \textbf{Type} \\ \hline
smoke weed everyday     & smoke                & weed                  & original      \\ \hline
snort cocaine everday   & snort                & cocaine               & synthetic     \\ \hline
inject heroin everday   & inject               & heroin                & synthetic     \\ \hline
chew mushrooms everyday & chew                 & mushrooms             & synthetic     \\ \hline
ingest drugs everyday   & ingest               & drugs                 & synthetic     \\ \hline
inhale nitrous everday  & inhale               & nitrous               & synthetic     \\ \hline
\end{tabular}}
\end{table}

\subsection{Data Quality}
The original dataset consisted of 2,661 tweets, each labelled as positive or negative. For verification of quality of labelling, a kappa test \cite{kappa} was performed to measure reliability. The labelling was performed by three graduate students (Punardeep Sikka, Zainab Kazi, and Mohiuddin Qudar); then a kappa statistic was generated against the original labelling to determine the consistency that the raters agreed on the label. As there were multiple raters a Fleiss' kappa \cite{fleiss1971mns} was performed and resulted in $0.6333$, indicating substantial agreement across all raters. It should be noted that this was executed to verify the quality of the original labelling, which had a field expert perform an assessment. The lower result can be attributed to a lack of knowledge of certain less common slang terms among the raters.

The resulting dataset may be considered small; however, falls in line with other similarly published works \cite{ma-etal-2018-rumor}. The set; however, did have an imbalance with 372 positive and 2,289 negative tweets. As such, synthetic generation proved necessary to combat this imbalance. Following the synthetic generation process mentioned in the previous section, the newly formed training dataset contained 6,790 drug-negative and 5,352 drug-positive tweets. Section \ref{results} further verifies the usefulness of explicitly labelled data from synthetic generation in improving the network's classification capability. Additionally, of the original 372 drug-positive tweets, only 150 were labelled as ``possibly\_sensitive" by the Twitter API. This initial result adds weight to the assumption that there is an inaccuracy in utilizing the tag for specifically detecting and classifying drug-related through the Twitter API.

\subsection{Support Vector Machine and XGBoost}
Support vector machines (SVM) are widely used in classification problems. However, when the dimension of the input data is large, SVM's are inefficient and take a great deal of time to train. To leverage this tool in our problem, we first used principal component analysis (PCA) to reduce the dimension of our word2vec model. Then we used the same data pre-processing method used for our CNN models to generate training data. We then trained the SVM on the testing data with different word vector dimensions. Since the difference of performance is not apparent for different word vector dimension settings, we chose 100 as the word vector dimension. The results are summarized in the confusion matrix shown in Table~\ref{confuse_svm}. Extreme gradient boosting (XGBoost) is a scalable tree boosting machine learning algorithm which supports parallel computing \cite{chen2016xgboost}. We used the same data for the SVM in XGBoost. The confusion matrix shown in Table~\ref{confuse_xgb} summarizes the results.

\begin{table}[h!]
\centering
\caption{SVM Confusion Matrix.}
\resizebox{0.7\textwidth}{!}{%
    \begin{tabular}{l|l|l|l}
\cline{2-3}
                                                 & \textbf{Labelled Correctly (True)} & \textbf{Mislabelled (False)} &                                 \\ \hline
\multicolumn{1}{|l|}{\textbf{Drug-Positive (1)}} & 117                            & 416                                & \multicolumn{1}{l|}{533}        \\ \hline
\multicolumn{1}{|l|}{\textbf{Drug-Negative (0)}} & 500                            & 8                                  & \multicolumn{1}{l|}{508}        \\ \hline
                                                 & 617                            & 424                                & \multicolumn{1}{l|}{\textit} \\ \cline{2-4} 
\end{tabular}%
}
\label{confuse_svm}
\end{table}

\begin{table}[h!]
\centering
\caption{XGBoost Confusion Matrix.}
\resizebox{0.7\textwidth}{!}{%
    \begin{tabular}{l|l|l|l}
\cline{2-3}
                                                 & \textbf{Labelled Correctly (True)} & \textbf{Mislabelled (False)} &                                 \\ \hline
\multicolumn{1}{|l|}{\textbf{Drug-Positive (1)}} & 78                            & 455                                & \multicolumn{1}{l|}{533}        \\ \hline
\multicolumn{1}{|l|}{\textbf{Drug-Negative (0)}} & 493                           & 15                                 & \multicolumn{1}{l|}{508}        \\ \hline
                                                 & 571                           & 470                                & \multicolumn{1}{l|}{\textit} \\ \cline{2-4} 
\end{tabular}%
}
\label{confuse_xgb}
\end{table}

\subsection{CNN-Based Classifier}
\subsubsection{Input Embedding}
\begin{figure}[h!]
  \centering
  \includegraphics[width=0.6\textwidth]{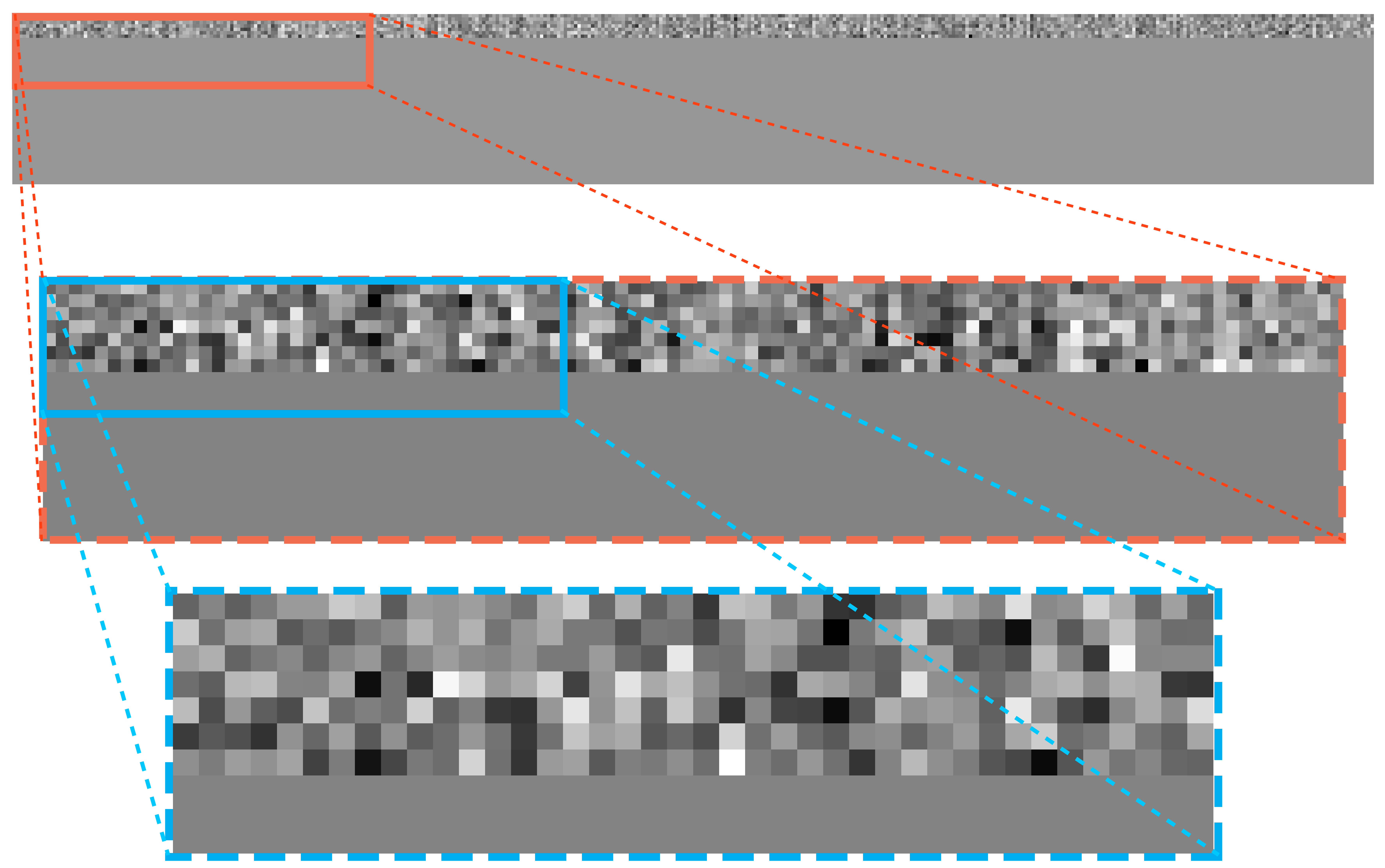}
  \caption{Visualization of the word2vec embedded text ``i think all drugs should be illegal". The first sub-illustration is the original word2vec embedded text ($50 \times 400$), the second and the third sub-illustrations are zoomed parts of the original embedding.}
  \label{w2v_demo}
\end{figure}

The individual inputs to the CNN were a fixed-size 2-dimensional embedded tweet text ($50 \times 400$) and each row in the input was a word2vec embedding. The order of rows in the input was correspondent to the order of words in the original tweet text. Unlike many formal English texts, tweets contain many misspellings. If a word2vec model trained on a structured corpus, such as Wikipedia or Google News, was used to embed the words from the Twitter texts, then there would have been a serious out-of-vocabulary (OOV) issue. In this scenario, if a word does not exist in said corpus then it cannot be embedded to the semantic-related vectors. Therefore, a word2vec model (referred to in this case as a Twitter word2vec model) pre-trained on a Twitter corpus \cite{godin2015multimedia} was utilized to embed the tweets. The dimension of embedding used in this paper was $K=400$, meaning each word in the word2vec vocabulary had a correspondent 400-dimensional unique vector. Based on the limitation of characters in each tweet (280 characters), the set length (number of words) of each input text was $L=50$. If the number of words in a tweet text was less than 50, then randomized vectors were applied (obeyed uniform distribution, the range was from -0.5 to 0.5) to extend the length of embedded text to 50. For a tweet text that had a length greater than 50, a sliding window of length 50 was employed to get the parts of the text. If any part of the text was labeled as positive (drug-related) by the CNN, a positive label was assigned to the whole text. Figure~\ref{w2v_demo} is a visualized 2-dimensional embedded text example. 

\subsubsection{CNN Architecture}
\begin{figure}[h!]
  \centering
  \includegraphics[width=0.65\linewidth]{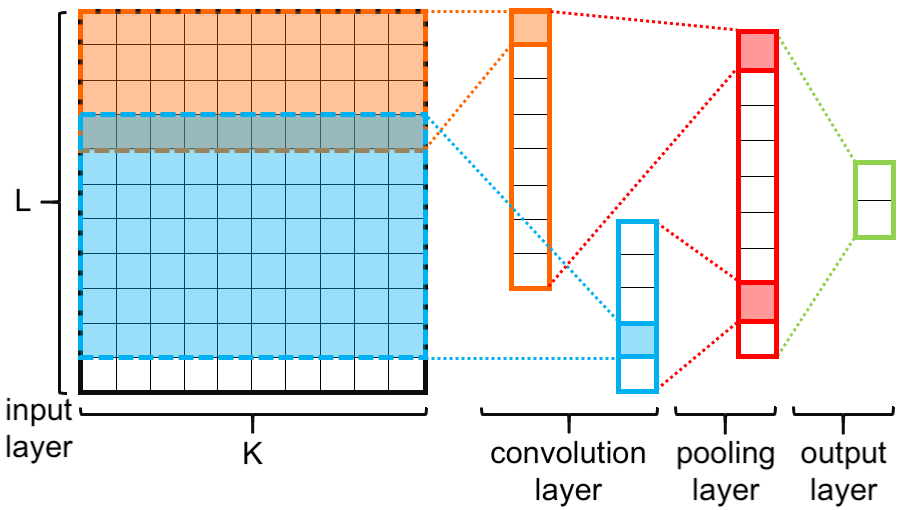}
  \caption{The CNN architecture. L is the length limitation of each sentence; K is the dimension of each Word2Vec embedding. Only two filters (orange and blue rectangles surrounded by dashed lines) are shown in this figure; therefore, only two feature maps shown in the convolution layer.}
  \label{cnn_arch}
\end{figure}

In this paper, a similar CNN architecture proposed by Kim \cite{kim2014convolutional} was applied. This CNN architecture had one convolution layer, and the length of each convolution filter was 400. The filters were; however, grouped by different heights. There were five groups of filters, where the respective height of each within the groups were 3, 4, 5, 6, and 7. Each group had 64 filters, so there were 320 in total. A one-max pooling approach was used to get the maximum values of each feature map and then concatenate them into an array of 320 values. The output layer, which as fully connected to the pooling layer, had two output neurons. Figure~\ref{cnn_arch} is a simplified version of the mentioned CNN architecture.

\subsubsection{Training}
The training batch size was 64 in the experiment. The Adam optimizer\cite{adamop} was used to minimize the loss value during training and the learning rate was set to $10^{-4}$. Since the number of positive labeled tweets in the training dataset was initially less than the number of negative labeled tweets, a weighted cross-entropy function was used as the loss function (see Equation.\ref{eq:loss}, where $Y$ represented the target labels; $\hat{Y}$ represented the predicted scores; and $\omega$ was the positive weight).

\begin{equation}
loss(Y, \hat{Y}) = Y[-log(\hat{Y})]\omega + (1-Y) [-log(1-\hat{Y})]
\label{eq:loss}
\end{equation}

The pseudo-code of the training algorithm is shown in Algorithm \ref{train-algorithm}. $T_{batch}$ is used to represent the set of tweets in a training batch (there is no overlap between any pair of batches and the union of all the training batches is the training tweet dataset $T$); $Y_{batch}$ is the set of labels of the corresponding training batch ($Y$ is the set of labels of $T$). $E_{batch}$ represents the word2vec embedded $T_{batch}$. $W2V$ represents the word2vec embedding dictionary. $\theta$ represents the parameters in the neural network model. $\gamma$ is the learning rate.

\begin{algorithm}[h!]
 \KwInput{The set of tweets: $T$; the set of labels: $Y$; the word2vec model: $W2V$}
 \KwOutput{A trained neural network parameters: $\theta$}
 Randomly initialize the parameters in $\theta$\;
 \While{$T$ is not empty}{
    Let $T_{batch}$ be the next batch of tweets from $T$\;
    Let $Y_{batch}$ be the corresponding set of labels from $Y$\;
    Remove $T_{batch}$ from $T$\;
    Let $E_{batch}$ be an empty set\;
    \While{$T_{batch}$ is not empty}{
        Get a tweet $tweet$ from $T_{batch}$\;
        Remove $tweet$ from $T_{batch}$\;
        Let $words$ be a list of the sequence of words in $tweet$\;  
        Let $embedding$ be an empty list\;
        \While{$words$ is not empty}{
            Get the first word $w$ from $words$\;
            Remove $w$ from $words$\;
            Let $w2v$ be the embedding of $w$\;
            \eIf{$w$ is in $W2V$}{
                Assign $w2v$ with output from $W2V$ model\; 
            }{
                Assign $w2v$ with random values between $-0.5$ and $0.5$\; 
            }
            Append $w2v$ to $embedding$\;
        }
        Make $embedding$ a $[50\times400]$ matrix, pad with zeros if the length of $embedding$ is less than 50\; 
        Append $embedding$ to $E_{batch}$\;
    }
    Let $\hat{Y}_{batch}$ be the set of predicted labels of $E_{batch}$\;
    Compute the loss (according to Eq.\ref{eq:loss}): $batch\_loss \leftarrow loss(Y_{batch}, \hat{Y}_{batch})$\;
    Use AdamOptimizer to optimize $\theta$ regarding $batch\_loss$\;
 }
\caption{Training algorithm (one epoch).}
\label{train-algorithm}
\end{algorithm}

In the experiment, the CNN model was trained separately on two datasets. One training dataset was the original with 2,661 manually labeled samples (372 labeled positively), while the other training dataset was the original+synthetic dataset with 12,142 samples (5,352 positive labeled samples). The best model selected was trained on two datasets, referring to them respectively as CNN model-A (trained on the original+synthetic dataset) and CNN model-B (trained on the original dataset). The testing accuracy of CNN model-A was 82.31\% and CNN model-B was 76.35\%, making the synthetic model an improvement over the original.

\section{Results}
\label{results}
\subsection{Evaluation Metrics}
The following metrics were used to evaluate each model: accuracy, precision, specificity and recall (sensitivity),  F1 score, receiver operating characteristic (ROC), and area under the curve (AUC). Although accuracy is the primary indicator of classification when the number of positive and negative testing samples are balanced, it is not comprehensive. A higher precision represents less false-positive predictions occurring, while a higher recall represents less false-negative predictions occur. F1 score is a function of precision and recall which is defined as $(2\times precision \times recall)/(precision + recall)$. We used the F1 score to measure the balance between precision and recall. AUC is derived regarding the ROC curve, which indicates the capability of a model to distinguish between classes. A high AUC value shows that corresponding model has a good distinguishing capability.

\subsection{Classification by Support Vector Machine and XGBoost}
\begin{figure}[h!]
  \centering
  \includegraphics[width=0.60\linewidth]{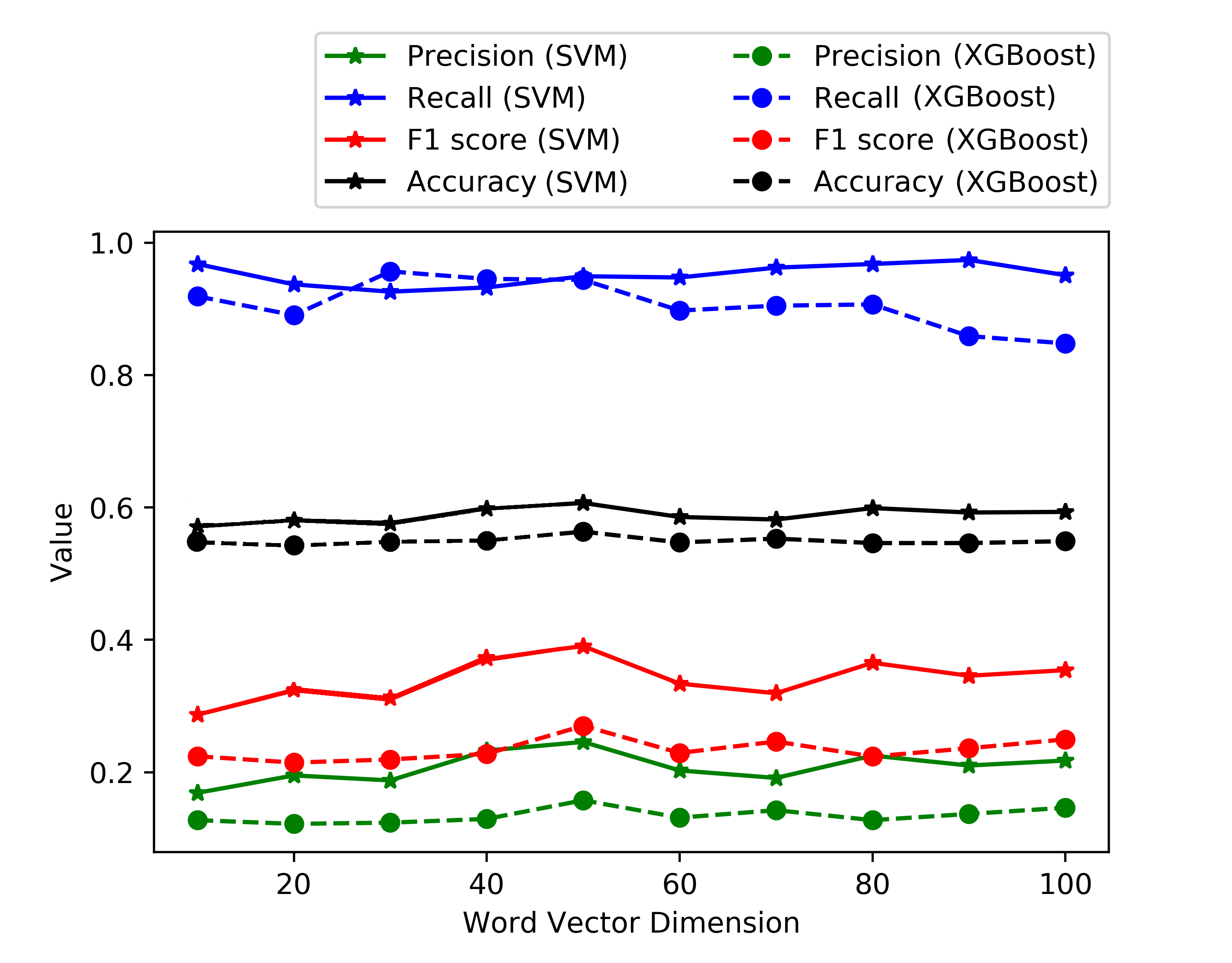}
  \caption{The testing precision, recall, F1 score, and accuracy of the SVM and XGBoost trained on the data with different word vector dimensions.}
  \label{svm}
\end{figure}

As mentioned previously, training SVM and XGBoost on the original word2vec-embedded tweets does not produce accurate results. Therefore, we used principal component analysis (PCA) to reduce the dimension of the word vectors before embedding the tweets. To compare the effect of different dimension reduction strength, we reduced the dimension of word vectors to different values (between 10 and 100), see Figure~\ref{svm}. The results indicate that there is no significant correlation between the PCA dimension reduction strength and model performance. Both SVM and XGBoost achieved relatively high recall, but low precision. Although the AUC of SVM appears to be satisfactory, the overall accuracy is relatively low. Therefore, we cannot consider the SVM and XGBoost as good models in this experiment.

\subsection{Classification by CNN-Based Classifier}
\label{cnn_results}
\begin{figure}[h!]
  \centering
  \includegraphics[width=0.6\linewidth]{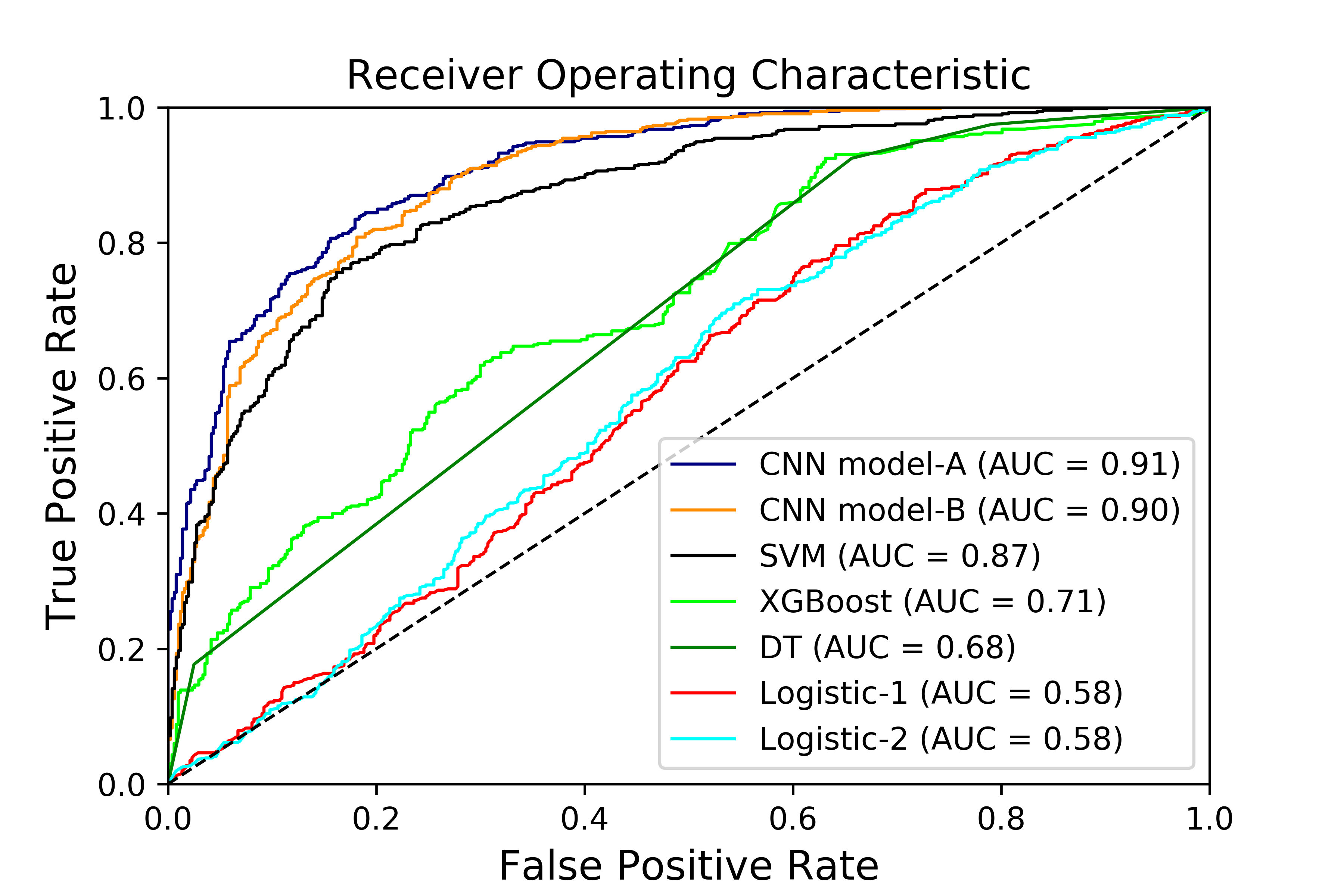}
  \caption{ROC curves for the  different machine learning models.}
  \label{fig:ROCs}
\end{figure}

In terms of AUC and accuracy, the neural network outperformed the previous models. Results utilizing decision trees and regression based models were not accurate enough for classifying this particular dataset. Instead the very structure and semantic meaning needed to be explored for significant conclusions. The results of all the models are summarized in Figure~\ref{fig:ROCs} and Table~\ref{fig:accuracies}. All classifiers were trained on a similar sized dataset in order to be properly compared against the neural network-based classifier. 

\begin{table}[h!]
\centering
\setlength{\tabcolsep}{3pt}
\caption{Metrics for the Different Machine Learning Models}
\label{fig:accuracies}
\resizebox{0.5\textwidth}{!}{%
    \begin{tabular}{|l|l|l|l|l|} \hline
     \textbf{Models} & \textbf{Accuracy} & \textbf{Precision} & \textbf{Recall} & \textbf{F1 Score}\\ \hline
     CNN model-A &  \textbf{82.31\%} & 0.893 & 0.784 & \textbf{0.835}  \\ \hline
     CNN model-B &  76.35\% & 0.597 & 0.906 & 0.719 \\ \hline
     Decision Tree &  63.40\% & 0.925 & 0.584 & 0.716 \\ \hline
     SVM &  59.33\% & 0.220 & \textbf{0.943} & 0.356 \\ \hline
     XGBoost &  54.90\% & 0.1463 & 0.8478 & 0.2469 \\ \hline
     Logistic-1 &  57.44\% & 0.873 & 0.546 & 0.672 \\ \hline
     Logistic-2 &  54.56\% & \textbf{0.954} & 0.525 & 0.677 \\ \hline
    \end{tabular}
}
\end{table}

Two CNNs were developed, and it should be of no surprise that the model trained with a larger set had an improved accuracy and AUC. However, a fascinating conclusion was the improved result originating from the synthetic data. While, having a similar AUC, the accuracy of the two models differs by a fair amount. This indicates that the synthetic data had a positive impact on the training of the network. It is possible that this can be attributed to expanding the results, considering keywords that may have been missed in the original training set. Although the ROC itself did not shift dramatically, diversifying the set further would most likely cause a worse classification in CNN model-B. The reasoning is simply that the model does not consider enough keywords, hence why the accuracy is lower. Regardless, the CNN proved to be the best classifier in terms of classifying drug-related tweets.

\begin{figure}[h!]
  \centering
  \includegraphics[width=0.7\textwidth]{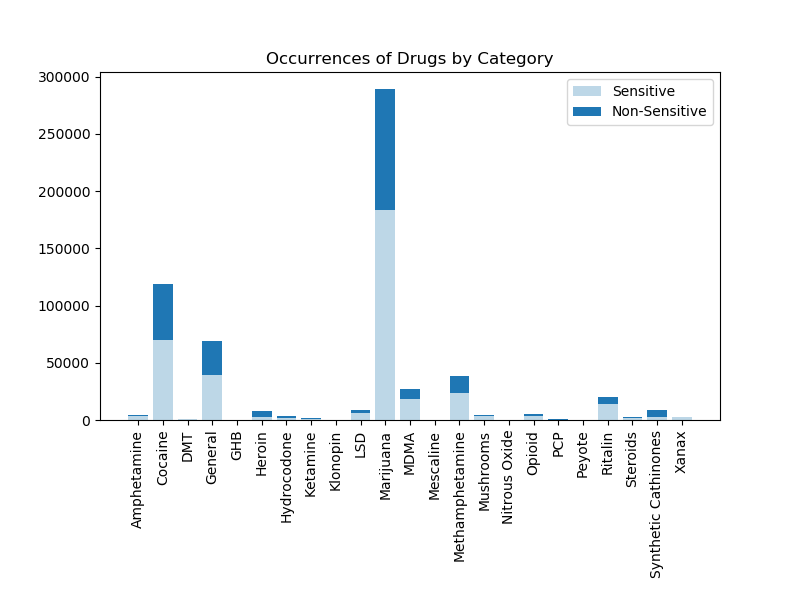}
  \caption{Specific Drugs Referenced by Category. The split in the bars correspond with classification made by the ``possibly\_sensitive" tag.}
  \label{fig:drug_occurences}
\end{figure}

\begin{figure}[h!]
  \centering
  \includegraphics[width=0.7\textwidth]{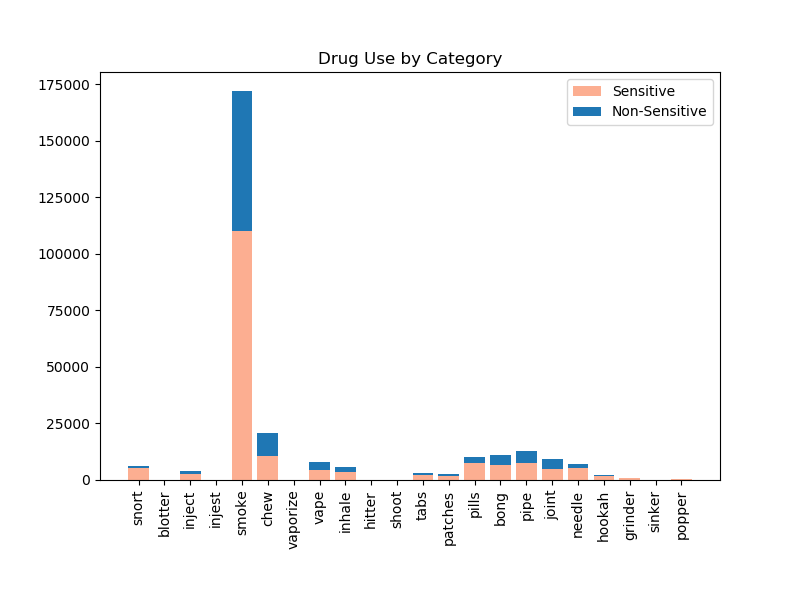}
  \caption{Drug-Use References by Category. The split in the bars correspond with classification made by the ``possibly\_sensitive" tag.}
  \label{fig:use_occurences}
\end{figure}

\begin{figure}[h!]
  \centering
  \includegraphics[width=0.7\textwidth]{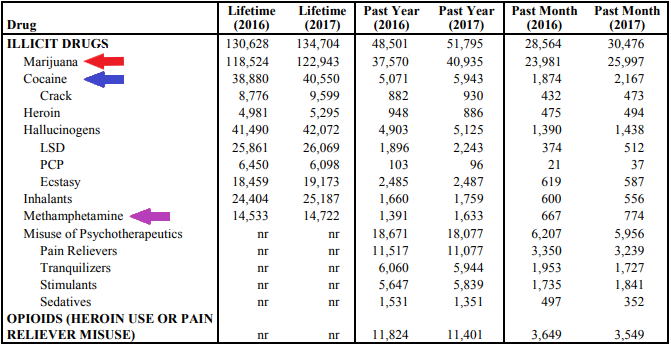}
  \caption{2017 Illicit Drug Use Survey Results\cite{drugsurvey}.}
  \label{fig:drug_society}
\end{figure}

Initial analysis on a subset of all the drug-positive tweets classified by the CNN (794,547 of the 3,696,150) was performed. As previously mentioned, the drugs were broken down into categories (by keyword). During pre-processing, the sums of each of these categories was taken based on the specific occurrence of a keyword within the tweet. The purpose of this exercise was to best determine the drugs referred to most frequently, as well as the occurrence of individual drug-uses. The results of this analysis can be viewed in Figures~\ref{fig:drug_occurences} and \ref{fig:use_occurences} respectively. As seen, the most common reference by a large margin was towards marijuana, with cocaine being a much lower second. An interesting facet of these numbers is they almost correspond with the literal drug activity displayed in true society \cite{drugsurvey}. Figure~\ref{fig:drug_society} was taken from a national survey, quantifying the most actively used illicit substances. If this is compared against the results in Figure~\ref{fig:drug_occurences}, then it can be seen that many of the common street drugs used were also the most commonly mentioned. The same idea can be said towards drug-uses, as smoking is the primary intake medium for marijuana. Though not all drugs align perfectly, many drugs such as marijuana, cocaine, and methamphetamine are still close.

Currently, Twitter utilizes a tag known as ``possibly\_sensitive" for declaring data that may be considered inappropriate to some readers. However, the tag does not specifically state why a tweet is targeted or if it can be utilized in classifying drug-positive tweets. Figures~\ref{fig:drug_occurences} and \ref{fig:use_occurences} show that the tag most likely cannot be used for this task. While drug-related tweets appear to be considered more sensitive than not, there are still a significant number that are missed. Therefore, one can most likely deduce that the ``possibly\_sensitive" tag is not a replacement for the classification done by the neural network.

\subsection{Additional Analysis}
\subsubsection{Keyword Strength Determination}
\begin{figure}[h!]
  \centering
  \includegraphics[width=0.8\textwidth]{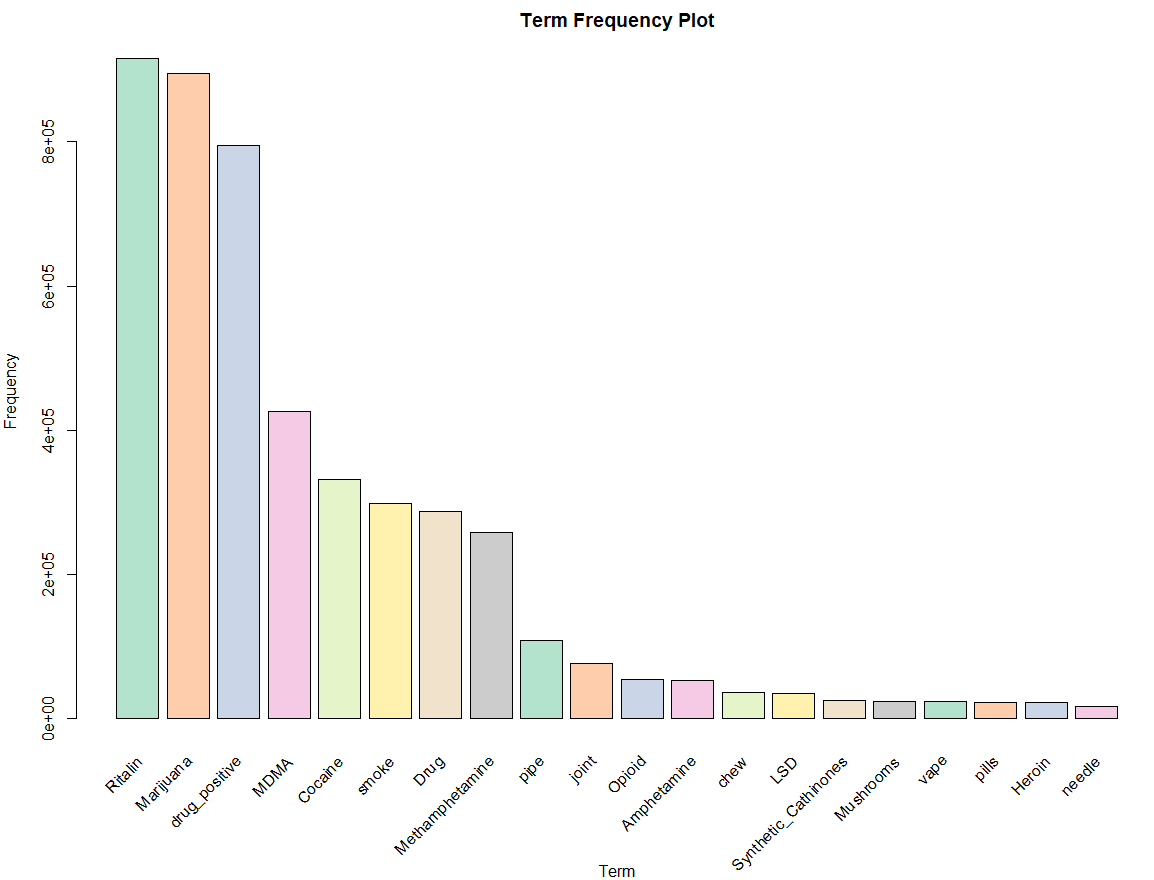}
  \caption{Tag frequency in the CNN classified data.}
  \label{fig:tagFreq}
\end{figure}

Following a similar process to that in \cite{Belyi2016}; after the data had been classified with the CNN, association rule mining was performed. The purpose of this process was to best determine the important relationships within the data. To begin the tags ``possibly\_sensitive" and ``drug\_negative" were removed from the data. The former meant that the tweet was deemed by twitter as a possibly sensitive tweet. The latter was classified by the CNN as having a negative association to drugs respectively. These tags were pruned as they clouded important rules within the data. The ``possibly\_sensitive" tag is present on every tweet that was put through the CNN, while the ``drug\_negative" is the opposite of the ``drug\_positive" tag. The balance between the two being 78.503\% and 21.497\% of the 3,696,150 tweets. The remaining tags that were considered in the association rule mining are ``drug\_positive" and the parent terms, meaning a term such as ``weed" would be considered as ``marijuana" in this example. In Table~\ref{tab:tagStats} the statistics that describe the amount of tags per tweet are shown. Frequency of specific tags is shown in Figure~\ref{fig:tagFreq}.  

\begin{figure}[h!]
  \centering
  \includegraphics[width=0.8\textwidth]{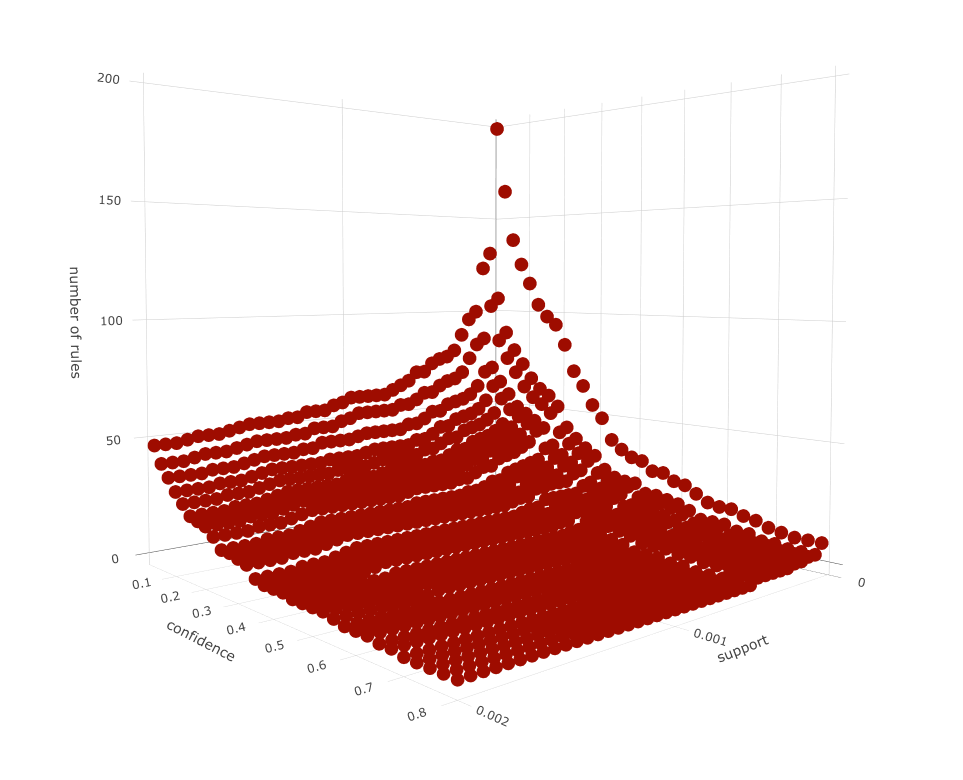}
  \caption{Association rule generation sensitivity, based on minimum support and minimum confidence.}
  \label{fig:sensitivity}
\end{figure}

\begin{figure}[h!]
  \centering
  \includegraphics[width=0.8\textwidth]{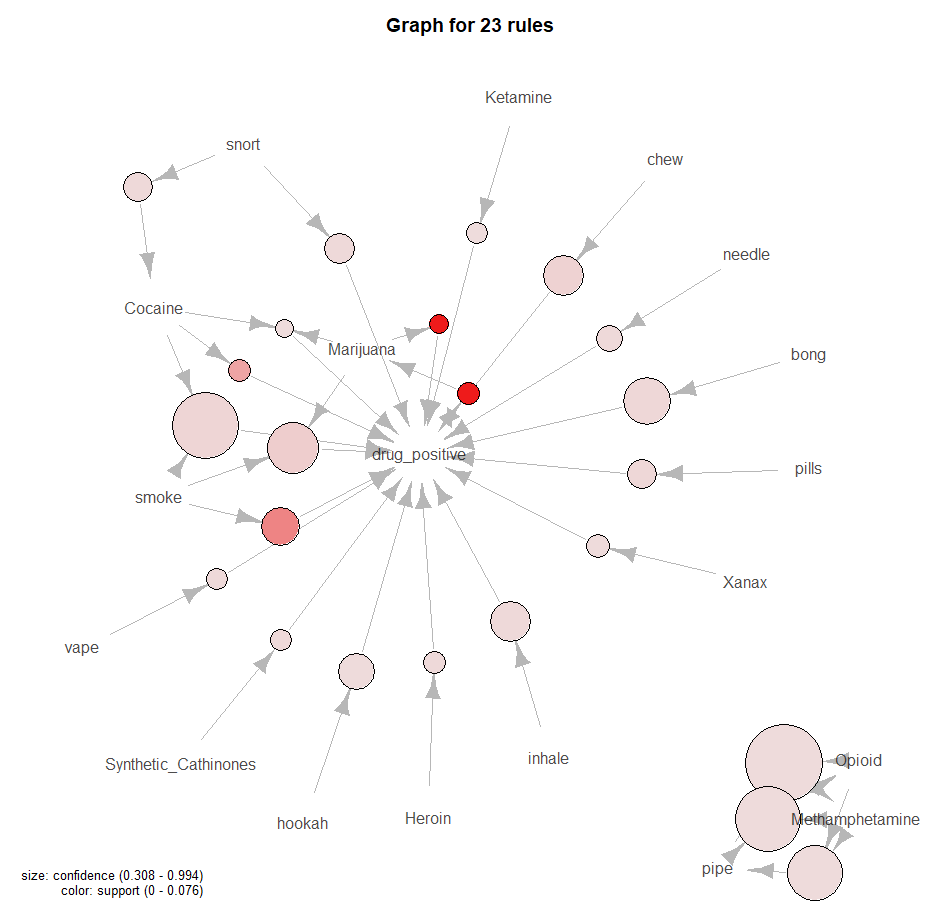}
  \caption{Network representation of generated association rules. Arrows from tag to circle is the antecedent of a rule and from circle to tag is the consequent.}
  \label{fig:network}
\end{figure}

A maximum of 5 tags was chosen for the mining of association rules, as it only excluded sets found in 80 tweets. Sensitivity analysis was then performed for the number of rules generated depending on the minimum support and confidence settings. This sensitivity can be seen in Figure~\ref{fig:sensitivity} from which a minimum support of 0.0003 and minimum confidence of 0.3 were chosen. The top 5 generated rules for the chosen minimum confidence and support can be seen in Table~\ref{tab:mined_rules}, where they are sorted by confidence. Full statistics for the 23 rules generated can be seen in Table~\ref{tab:mined_rulesStats}. A network representation of the rules can be seen in Figure~\ref{fig:network}. We applied the HITS algorithm \cite{10.1145/511446.511514} that is designed for finding hubs and authorities in the context of websites to the rule set. The non-zero hubs in order of precedence were found to be ``methamphetamine", ``pipe" and ``opioid". The non-zero authorities in order of precedence were ``opioid", ``pipe" and ``methamphetamine". This shows that there appears to be no relationship between the rules that have ``drug\_positive" in the consequent, meaning they independently hold. This is due to the fact that the only ``loop" present in the portion of the network containing the rest of the rules is between ``marijuana" and ``drug\_positive". Examining Table \ref{tab:mined_rules}, we can see that the strongest relationships revolve around largely the trio ``opioid", ``pipe" and ``methamphetamine". Since this is lacking ``drug\_positive", we can conclude that discussion around the topics of opioids and methamphetamine is likely part public concern about the problem. Relations between ``marijuana" and ``cocaine" in regards to ``drug\_positive", indicates that these are the most frequent drugs that twitter users partake in, and are willing to discuss on the platform. An extension to this is that ``smoke" is involved in both of these relations indicating that it is the preferred vehicle for delivery of these substances.

\begin{table}[h!]
\centering
\setlength{\tabcolsep}{3pt}
\caption{Statistics that describe the amount of tags per tweet.}
\label{tab:tagStats}
\resizebox{0.35\textwidth}{!}{%
    \begin{tabular}{|c|c|} \hline
     Statistic & Number of Tags in Tweet \\ \hline
     min & 1.0\\ \hline
     1st Qu. & 1.0\\ \hline
     Median & 1.0\\ \hline
     Mean & 1.309\\ \hline
     3rd Qu. & 2.0\\ \hline
     Max & 11.0\\ \hline
    \end{tabular}
}
\end{table}

\begin{table}[h!]
\centering
\setlength{\tabcolsep}{3pt}
\caption{Statistics for mined association rules.}
\label{tab:mined_rulesStats}
\resizebox{0.45\textwidth}{!}{%
    \begin{tabular}{|c|c|c|c|c|} \hline
     statistic & support & confidence & lift & count\\ \hline
     Min. & 0.0003728 & 0.3081 & 1.433 & 1378  \\ \hline
     1st Qu. & 0.0006541 & 0.3482 & 1.591 & 2418  \\ \hline
     Median & 0.0020700 & 0.4359 & 2.063 & 7651  \\ \hline
     Mean & 0.0116810 & 0.5090 & 6.202 & 43175 \\ \hline
     3rd Qu. & 0.0048150 & 0.6037 & 3.096 & 17797 \\ \hline
     Max. & 0.0762532 & 0.9945 & 58.176 & 281843 \\ \hline
    \end{tabular}
}
\end{table}

\begin{table}[h!]
\centering
\setlength{\tabcolsep}{3pt}
\caption{Top 5 mined association rules by confidence.}
\label{tab:mined_rules}
\resizebox{0.75\textwidth}{!}{%
    \begin{tabular}{|c|c|c|c|c|} \hline
     rule & support & confidence & lift & count\\ \hline
     opioid, pipe $\rightarrow$ methamphetamine & 0.0005367749 & 0.9944862 & 14.230069 & 1984  \\ \hline
     cocaine, smoke $\rightarrow$ drug\_positive & 0.0041372791 & 0.8685182 & 4.040260 & 15292  \\ \hline
     methamphetamine, pipe $\rightarrow$ opioid & 0.0005367749 & 0.8562797 & 58.175785 & 1984 \\ \hline
     methamphetamine, opioid $\rightarrow$ pipe & 0.0005367749 & 0.7342709 & 25.161546 & 1984 \\ \hline
     marijuana, smoke $\rightarrow$ drug\_positive & 0.0089966611 & 0.6865773 & 3.193889 & 33253 \\ \hline
    \end{tabular}
}
\end{table}

\section{Conclusion}
\label{conclusion}
The purpose of this work was to classify drug-related tweets and extract feature related information from the results. We achieved a high classification rate while implementing a novel methodology in adding synthetic data to the training process. As it stands, the results have shown that simple analysis without the CNN is limited and possibly fruitless. Regarding the CNN, as there are an infinite number of possible misspellings, the OOV problem cannot be eliminated. However, a character-level CNN \cite{zhang2015character} can be used to match some of the OOV words to the most similar words in the word2vec vocabulary, thus reducing the amount of OOV words. Furthermore, as manually labeling drug-related training data is very expensive, the neural network model can be pre-trained on some large and easy-to-reach dataset (such as the Twitter dataset for negative/positive sentiment classification). A smaller learning rate can then be implemented to train the neural network model on the current dataset. 

It should be noted there is a limitation with the current model, which is due to a lack of diversified data and a constraint from physical labelling. The model was trained for a particular period when certain drugs were more prevalent than others. As an example, the Canadian government had legalized Marijuana during this period, making is highly discussed topic. More training data will be needed to deal with such issues, and possible changes in the way people post may need to be considered in the future. Regardless, the model was accurate given the input set and even matched actual drug use trends in society. This may exhibit the possibility of using social media more often as a surveying tool, obtaining metrics and real behavioral trends over performing separate manual measurements.

Subword embedding was not considered for this work \cite{Bojanowski2016EnrichingWV} and will most likely prove useful in situations where the morphological structures of a particular term hold importance. A such, this methodology will be considered in future work to further improve the model. Additionally, due to the exponential growth of social media data, real-time data processing is essential in practice \cite{shahframework}. Providing solutions to the challenges such as dynamic updates in the training dataset and the filtration of spam tweets \cite{robinson2018birds} is the next step to take.

\clearpage
\section*{Competing interests}
  The authors declare that they have no competing interests.
  
\section*{Research Ethics Approval}
The Principal Investigator has the Lakehead University Research Ethics Board (REB) to conduct this research.

\section*{Author's contributions}
\emph{Conceptualization}: Joseph Tassone, Peizhi Yan, Mackenzie Simpson.\\
\emph{Data curation}: Chetan Mendhe, Joseph Tassone.\\
\emph{Formal analysis}: Joseph Tassone, Peizhi Yan, Mackenzie Simpson.\\
\emph{Methodology}: Joseph Tassone, Peizhi Yan, Mackenzie Simpson.\\ 
\emph{Supervision}: Vijay Mago, Salimur Choudhury.\\
\emph{Visualization}: Joseph Tassone, Peizhi Yan, Mackenzie Simpson.\\ 
\emph{Writing – original draft}: Joseph Tassone, Peizhi Yan, Mackenzie Simpson.\\
\emph{Writing – review \& editing}: Joseph Tassone, Vijay Mago, Salimur Choudhury.

\section*{Acknowledgements}
The authors would like to sincerely thank Punardeep Sikka, Zainab Kazi, Mohiuddin Qudar, Mannila Sandhu, and Tanvi Barot for their time in manually labelling the test and training samples employed by the CNN. 
Additionally, they would like recognize Longfei Zeng, Dillon Small, and the overall Lakehead DaTALab for providing support and the initial dataset. Lastly, they would like to acknowledge Caleb Pears for acting as a substance and addictions consultant, and verifying the keyword selection.


\bibliographystyle{unsrt}  
\bibliography{references}

\end{document}